\documentclass[11pt,twocolumn]{scrartcl}

\usepackage{casa_conf}	
\usepackage{graphicx}	

\title{Immersive Technologies and Elderly Users: Current use, Limitations and Future Perspectives}

\author {Zoe Anastasiadou\\ Visual Media Computing Lab\\
 Cyprus University of Technology\\ Limassol, Cyprus\\zx.anastasiadou@edu.cut.ac.cy
\and 
Andreas Lanitis\\Visual Media Computing Lab\\
 Cyprus University of Technology\\ Limassol, Cyprus\\andreas.lanitis@cut.ac.cy}

\begin{document}

\maketitle

\begin{abstract}
The increase of the percentage of elderly population in modern societies dictates the use of emerging technologies as a means of supporting elder members of the society. Within this scope, Extended Reality (XR) technologies pose as a promising technology for improving the daily lives of the elderly population. This paper presents a literature review that describes the most common characteristics of the physical and mental state of the elderly, allowing readers, and specifically XR developers, to understand the main difficulties faced by elderly users of extended reality applications so they can develop accessible, user friendly and engaging applications for the target audience. Furthermore, a review of existing extended reality applications that target the elder population is presented, allowing readers to get acquainted with existing design paradigms that can inspire future developments. 
\end{abstract}
\linebreak
\linebreak


\section{Introduction}
In recent years, Extended Reality (XR) has emerged as a promising technology in various fields, including healthcare, entertainment, and education. Among its many applications, XR holds great promise for improving the quality of life of older adults. With the global population aging, the need for innovative solutions that support the physical, mental, and emotional well-being of older adults is more urgent than ever. XR offers a unique way to engage older adults, providing them with opportunities for cognitive stimulation, physical rehabilitation, and social interaction. Its ability to create immersive and interactive experiences makes it particularly attractive to older users, who may face limitations in mobility, social engagement, or cognitive function due to aging   \cite{b1}. 
Currently, XR is being used in various healthcare settings to support elderly patients in rehabilitation and cognitive training. For example, XR-based exercise programs help older adults recover from strokes or injuries by guiding them through immersive, virtual environments that encourage movement and motor skill development   \cite{b2}. In addition, XR systems have been used in memory and problem-solving exercises to enhance cognitive function, especially for older adults at risk of developing dementia or Alzheimer's disease. These applications have shown encouraging results, offering a new and attractive approach to managing the health challenges that often come with aging. Furthermore, XR has also been used to combat social isolation, allowing older adults to interact with family members, friends, or even virtual environments such as museums or exotic locations   \cite{b3}. Despite its potential, the use of XR among older adults has several challenges. A major barrier is the physical and cognitive demands that VR places on users. For older adults, especially those with limited technological proficiency or age-related problems, using VR equipment can be complex and difficult. Headsets can be uncomfortable, and navigating virtual environments can be overwhelming for users, particularly those with reduced cognitive skills   \cite{b4}. In addition, issues such as nausea and dizziness are commonly reported by older adults, further hindering effectiveness.
This paper presents a literature review that describes the most common characteristics of the physical and mental state of the elderly, allowing readers, and specifically XR developers, to understand the main difficulties faced by elderly when using Virtual and Augmented Reality applications so that they can develop accessible, user friendly and engaging applications for the target audience. Furthermore, a review of existing Virtual and Augmented Reality applications that target the elder population is presented, allowing readers to get acquainted with existing design paradigms that can inspire future developments. Finally, future perspectives regarding the use of immersive technologies for this specific audience, along with concluding comments are presented.

\section{Characteristics of Elderly}
There is no agreement on when a person becomes old. Chronological boundaries, such as chronological and biological age, initially identify life-cycle changes and the principle of old age, but in essence, the principle of old age is marked by a combination of chronological, functional and social parameters. The UN accepts the age of 60+ as an adult enters old age   \cite{b5}. The age of 65, which coincides with the retirement age in most developed countries, is considered to be the beginning of old age \cite{b6}. Seniors define successful aging as an active method that includes events and exercises that focus on promoting or keeping physical, functional, psychological, and social health \cite{b7}. According to world epidemiological data \cite{b8} soon older people will be more than children for the first time in the history of humanity.
In general three subgroups of elderly are distinguished: The first consists of so-called young adults, aged 60+ or 65+ up to 74 years old (young old). These people are usually fit, active, and take care of themselves. They present quantitatively the least medical and social problems. The second subgroup relates to people aged 75+ to 84 years old (mid old). After the age of 75 the condition of the elderly is becoming more fragile and is characterized by severe physical and mental impairment. Finally, the third subgroup consists of older people over 85 years (the oldest old group) of age who are usually very fragile, have severe physical and mental impairments, are highly dependent on other people, and due to mobility problems are usually confined to home. 
Technology has the potential to enhance the lives of older adults by improving their safety, security, and self-confidence in everyday life. However, too often older adults' abilities and limitations are not reflected in the design of current and future technologies. Although older adults experience specific limitations as they age, the word old does not necessarily identify people who are disabled or sick. Many people over the age of 65 are reportedly in good health \cite{b9}. The involvement in end-of-life education promotes successful aging, as the activity of learning in old age can positively influence the quality of life and overall well-being \cite{b10}.
Physical activity is seen to present an important role in promoting the functional status, psychological status, and well-being, and has social advantages \cite{b11}. The overall physical abilities and appearance of people deteriorates as they grow older \cite{b12}. In terms of vision, these include changes in visual acuity, color perception, and susceptibility to glare. In the auditory domain, older adults face higher difficulty perceiving high-pitched sounds and greater interference from background noise \cite{b13}. In a very literal sense, older adults may perceive technology differently than younger adults do \cite{b14}. 

\subsection{The effect of aging on the abilities of elders} 
Over time, physical and functional changes occur in humans that are not necessarily based on chronological age. These changes push people to appreciate their abilities and ways of dealing with those who interact with others and the natural environment \cite{b15}. In this section age related-changes are categorized in sensory, physical and cognitive changes, and a relevant discussion for each category is presented. It is important to note that good design for the elderly is often a good design for everyone so designers should take into account sensory, physical and cognitive factors to ensure the adoption of design for all principles. 
One of the first signs of aging is the weakening of the near focus or presbyopia, where it is repaired with reading glasses or binocular lenses. Most people first notice this weakening in their early 40s. The focus on the nearest point is 10 cm at the age of 20, contrasted to 100 cm at the age of 70. People over the age of 70 are more likely to have some degree of vision and hearing loss. Another feature of vision that changes over time is the field of view where there is a reduction due to factors such as the decline characterized by the eyelash tilt \cite{b16}. In late adulthood, the deterioration of visual acuity affects the eyes of the elderly where it is particularly evident in many diseases such as cataracts, macular degeneration, glaucoma, and diabetic retinopathy \cite{b17}. 
The understanding of touch, pressure, and vibration decreases with age, especially in the hands and feet \cite{b18}. The sense of touch is the creation of two subsystems. Sensation of the skin records information about vibrations, pressure, heat, cold and tissue damage (pain). The proprioception, the sense of the machine system, is able to determine the position and movement of the limbs and the collision of external forces \cite{b19}. Lower pressure sensitivity makes it harder to feel when the body is in full touch with a surface or when a small surface is pressed, such as an elevator button or keypad \cite{b20}. 
Elderly people show several physical changes including muscle weakness, vertebral compression fractures, and/or back muscle weakness are known to be related to kyphotic curvature of the thoracic and/or lumbar spine in elderly patients \cite{b21}. The kyphotic curvature of the spine negatively affects the quality of life, thus affecting the experience of using AR in the elderly. A study revealed that declines in physical and mental health, the loss of functional capabilities, and a weakening of social ties represent obstacles to active aging among institutionalized older adults \cite{b22}. In the elderly population the range of body motion is limited, muscle strength is reduced, the body is less flexible and reflexes are more moderate. The height of the trunk and the length of the arm are reduced, making it more difficult to access items. Furthermore, diseases, such as arthritis, cause joint hurt and stiffness, making it tough to grip and hold different surfaces\cite{b23}. 
Diseases that occur frequently in elders, such as Parkinson's, stroke or mechanical damage to the facial nerve, can lead to dysfunction of facial muscle movements. One consequence of this is that the structure of daily life has to adapt to the health issues. For example even simple tasks like food intake require longer time due to difficulties in eating and swallowing \cite{b24}.
Age-related changes in cognitive function vary considerably between individuals and cognitive domains, with some cognitive functions appearing more sensitive than others to the effects of aging. Much of the basic research on cognitive aging has focused on attention and memory, since deficiencies in these fundamental processes may explain much of the variation observed in higher-level cognitive processes \cite{b25}. Dementia is a term used to describe a decline in mental abilities, including memory, language, and logical thinking, that is severe enough to affect daily living \cite{b26}. An important element of the research is that older adults usually appear to activate different brain structures than younger adults when performing cognitive tasks \cite{b27}. 

\section{Difficulties faced by the elderly in Using XR}
Virtual Reality is defined as the use of computer technology to create the effect of an interactive three-dimensional world in which the objects have a sense of spatial presence \cite{b28}. Extended Reality (XR) is a broad term that includes augmented reality (AR), virtual reality (VR), and mixed reality (MR). AR blends virtual elements with the real world in real-time, while VR enables users to interact with and explore a simulated environment, whether it’s a replica of the real world or an entirely fictional one. These technologies are frequently combined to create more engaging and immersive experiences \cite{b29}.VR utilizes interactive computer-generated 3D environments \cite{b30}, primarily incorporating auditory and visual feedback, and occasionally haptic feedback. VR can be categorized into non-immersive, semi-immersive, and fully immersive systems \cite{b31}. Non-immersive setups are desktop-based with limited interaction (e.g., keyboard, joypad) and immersion (e.g., PC, tablet). Semi-immersive systems typically feature a large monitor/projector with moderate interaction and immersion (e.g., Kinect, data gloves). Immersive systems utilize tools like head-mounted displays (HMDs) or the cave automatic virtual environment (CAVE) for high interaction (e.g., trackers) and deep immersion in the virtual environment (VE). Moreover, VR can be viewed as a spectrum between reality and virtually, where aspects of the VE blend with the real world (augmented reality) or vice versa (augmented virtually) \cite{b32}.
Extended reality holds great promise as a tool that could improve the treatment of cognitive and emotional disorders in the elderly. While XR applications have been successfully used in clinical environments with adolescents and children, there has been comparatively less research conducted on its application in the geriatric population \cite{b33}. Some XR applications are designed to combine virtual representations with the perception of the physical world. Stereo sound may contribute to the immersion, as well \cite{b34}. Many extended reality applications are about combining real-world virtual representations providing the user with information about the natural world that they would not be able to obtain otherwise. This type of application is referred to as augmented reality. With this technology, the user perceives the real world by receiving information invoked by virtual elements so instead of experiencing the physical reality, one is placed in another reality that includes the physical along with the virtual \cite{b35}. Extended reality is typically associated with younger populations, as they are sophisticated technologies considered to be the most suitable for gaming platforms and have features such as complexity for creating and using them. XR technology has quickly been used in a number of cases, including games, navigation, medicine, education and design \cite{b36}.
XR can play an important role for maintaining for the wellbeing of older people, hence it is necessary to recognize the challenges and difficulties faced by older people when using XR technologies, and study the reasons that make elder people reluctant to utilize those technologies. In a community where our elderly population is increasing significantly, we have to find ways to support aged people.
As we mentioned above XR technology offers the user immersion into a virtual world where they can react and interact with the environment enabling them to have realistic experiences. Augmented reality enhances this situation by providing a realistic environment with additional virtual information. Currently, older adults are being exposed to technologies and improving their expertise as concerns using interactive technologies with the support of younger people \cite{b37}. For technology-based learning activities, a typical person needs perceptual, motoric, and cognitive skills. However, as people age, their cognitive, physical, and sensory skills tend to decline. Furthermore, the high cost for private use of such technologies hinders the accessibility of technological tools to elder people. Additionally, older adults may refrain from using the latest technologies because they believe that modern technologies are difficult to control or impossible to use.
Technologies such as XR have been identified as a solution used to assist elders who face various difficulties later in life. The principal purpose of these emerging technologies is to promote their care and control their health. Social and emotional well-being has also recently been recognized as an important factor in improving the quality of life of elder people. Though, there is a deficiency of XR technologies to support their social and emotional well-being. New technologies that have been designed without understanding the characteristics of older adults are limiting their experiences with these technologies or even harming helpless users. To answer the question of whether older people can use technologies by incorporating them into their daily lives, proposed an evaluation framework, focusing on three categories: physical, social, and psychological well-being. More details about each of these categories are presented below:
XR can independently meet the needs of older adults by recognizing and understanding their behavioral patterns and can make choices for users when needed. When this design principle is applied, it is possible to defeat the anxiety, worry, and stress that elder people feel while using the technology. Many studies on XR have centered on old adults with a disability. Most studies of interactions between older adults and XR use have defined the age of the user as a criterion for the successful use of the application. Age is not a predictor for use of interactive technology. It is a fact that physical capacity limits as one becomes old.
Older adults have different levels of skills and abilities and different tastes and needs and it is necessary to recognize this fact. XR technologies can be designed and adapted to the individual's cognitive level and movement issues. When this design principle is applied, it is possible to reduce the fear, difficulty, and stress they feel while using the new technology \cite{b38}.
Researchers \cite{b39} mentions an issue that arises for the design part of VR applications for people with dementia. Also focuses on the settings they need at the beginning of the application for the user, which requires technical assistance from caregivers or family members who should be able to use any application for support of people with dementia so they can use it later. There is a risk that designers will use the "One Size Fits All" method. Any interruption of assistive technology should be designed to reflect the necessities of the disease and should be considered the idea that their needs may vary over time depending on how the situation of their health affecting them at the specific moment.

\section{Overview of Existing XR Applications for Elderly Users}
There is a generally supported view of the interest of keeping older adults in sync with the latest technological developments. As well as joining with family and friends, technology can support older adults in improving social support, increasing access to medical knowledge, enabling them to engage as citizens in decision-making processes, and allow them to keep fit using dedicated fitness apps \cite{b40}.
In today's aging society, falls represent a significant public health issue for the elderly population. While VR technology shows promise in mitigating fall risk \cite{b41}, the lack of user body representation in VR environments diminishes spatial presence. Augmented reality, offering greater presence and embodiment, presents an alternative. To address this, Chen et al., \cite{b42} created an AR-based exergame system tailored for the elderly to reduce fall risk. Data analysis revealed positive user evaluation, with pragmatic quality rated as good and hedonic quality as excellent.
Stammler et al., \cite{b43} describe an AR-based application named Negami for spatial neglect treatment, integrating visual exploration training with active, eye, head, and trunk rotation. The app overlays a virtual origami bird into the patient's real environment (see figure 1), which they explore using a tablet camera. Subjective feedback from healthy elderly participants (n=10) and stroke patients with spatial neglect who used the Negami app was analyzed. Usability, side effects, and gaming experience were assessed through various questionnaires. Healthy elderly participants found training at the highest difficulty level to be challenging but not frustrating. The app received high ratings for usability, minimal side effects, and high levels of motivation and enjoyment. Stroke patients with spatial neglect consistently rated the app positively in terms of motivation, satisfaction, and enjoyment. 

XR has been employed in various areas for rehabilitation purposes, especially in stroke recovery. Consequently, recent guidelines have advocated for the integration of XR into both motor and cognitive rehabilitation programs for stroke survivors \cite{b44}. Peleg-Adler et al., \cite{b45} studied the practicality of AR technology for older adults by observing their interactions and performance in a path-planning task compared to younger adults. They assessed task completion time, error rate, device movements, and subjective impressions using both AR and non-AR interfaces. Forty-four participants from two age groups were selected: community-dwelling, healthy adults over 65, and younger individuals aged 25–40, mostly university students. Older adults were slower in both interfaces, consistent with age-related declines in perceptual and cognitive skills. However, the impact of AR on performance was similar across age groups. Despite being unfamiliar with AR, older adults showed similar performance changes as younger adults when using AR compared to non-AR applications. Interestingly, older adults preferred the AR interface and reported a better user experience compared to younger participants. They found the AR application user-friendly, practical, easy to learn, and intuitive. These findings suggest that older adults are willing and capable of embracing new technologies, especially when they perceive benefits in their daily lives.
More recently, Kosti et al., \cite{b46} indicate that seniors showed a strong interest in XR and modern technologies, with a desire to generate their own content. This content was integrated into the virtual village's outdoor features. Participants were motivated by the interactive and social aspects of the VR environment, feeling both creative and exploratory. Image sentiment analysis confirmed their positive reception, indicating high levels of acceptance. 
According to Rojo et al., \cite{b47} ensuring adherence to physical exercise training is crucial for older adults and individuals with neurological disorders. Immersive technologies, increasingly incorporated into neurorehabilitation therapies, offer potent motivational and stimulating elements. This study aims to assess the acceptance and potential safety, utility, and motivation of a developed VR system for pedaling exercise among these populations. A feasibility study was conducted with patients from Lescer Clinic and elderly individuals from the residential group Albertia. Participants engaged in a pedaling exercise session using the VR platform, PedaleoVR (see figure 2). The findings indicate that PedaleoVR is considered a credible, user-friendly, and motivating tool for adults with neuromotor disorders to perform cycling exercises, potentially enhancing adherence to lower limb training activities. Furthermore, PedaleoVR does not induce negative effects associated with cyber sickness, while the sense of presence and overall satisfaction among the geriatric population were positively evaluated.
With the right combination of technologies it can be very useful about the elderly, as for example artificial intelligence-based robots used for social care promote social exchange and emotional relaxation, and the use of exciting content can enhance life satisfaction and family ties \cite{b48}. Tammy Lin and Wu \cite{b49} study showed that the Proteus effect, which involves adjusting avatar age in VR, works well for older people when they exercise. The findings demonstrated that, for older individuals who did not participate in strenuous exercise, the VR embodiment of younger avatars causes a larger perceived exertion of exercise.

\section {XR and Dementia}
Elderly people often treasure their memories of the past, and by sharing them, they can gain a sense of self-existence; however, due to the gradual decline of memory, all that remains are deep memories of the past; as a result, they tend to vividly repeat the same matter, but frequently forget more recent things and people \cite{b50}. XR applications could help persons with dementia \cite{b51}. The mix of interactivity and the ability to give high-fidelity 3D visualizations of places and objects provides viewers with a one-of-a-kind experience that is impossible to replicate with traditional media. For example, virtual worlds could be used as memory aids, allowing users to engage with artifacts that are difficult or expensive to obtain in the real world, or to provide an experience of a location that is no longer accessible, such as a famous street from the past \cite{b52}. The interactivity of virtual worlds may enable older individuals to participate actively in a more playful and creative recollection experience, rather than just being passive observers of things and settings.
Narrative gerontology is known for enhancing the well-being of elders through intergenerational sharing, promoting "well-aging." Digital storytelling, utilizing sound, images, and music, can convey a biography in minutes, sharing it with a small circle or a global audience. By merging narrative gerontology with digital storytelling into "digital narrative gerontology," we aim to introduce a new concept that will positively impact elders' well-being and aging process, while also sharing these beneficial values with the broader community \cite{b53}. 
Wu et al., \cite{b54} describe an experiment with 12 patients with mild dementia. The experiment included observations of behavior for three weeks during memory therapy activities. In the preliminary stage of this study, the existing digital narration application software is first analyzed through the most popular applications such as Storybirds (https://www.storybird.com/), Photo Story (https://shorturl.at/mNP09), Puppet Pals (https://shorturl.at/bgU37) and Toontastic Jr. Pirates (https://shorturl.at/buvP6). A questionnaire survey was then conducted in Old Taiwan which is an old country as the elderly remember it, that corresponded to the treatment of memory images, for the image database of the application mentioned below. There were a total of 185 questionnaires were conducted in categories, including Taiwanese characteristics, childhood, food, costume, architecture, vehicle and education, etc. As part of this effort, the "ReStor" app was created, which allows them to maintain a tale intact while presenting it through the app and sharing it with others without fear of forgetting or repeating it. The "ReStor" design mixes recordings and graphics to communicate narrative information, making the story production process more intuitive and decreasing interface operation difficulties, which can effectively assist the elderly with dementia in tale creation. Memory will be enhanced by the re-narration of earlier stories, which will help to halt the progression of dementia. It is envisaged that the "ReStor" technique will help the elderly retrieve their memory and cognitive abilities by reopening earlier stories. 
Siriaraya and Ang \cite{b55} aimed to explore the use of 3D virtual world technology to engage individuals with dementia in long-term care. They developed three versions of virtual world prototypes (reminiscence chamber, virtual tour, gardening) using gesture-based interaction, focusing on older dementia patients and their caregivers. The prototypes utilized gesture and touch-based interfaces with Microsoft Kinect sensor and ZDK middleware in Unity3D, displayed via projector. Initially designed for whole body interaction, later versions simplified to arm movements due to resident fatigue. Prototypes included a virtual room with historical artifacts and music, virtual tours, and a virtual garden. Data collected through observations highlighted the potential of virtual worlds to stimulate memories and improve self-awareness. The study emphasized the importance of considering design issues and the complementary role of virtual and physical activities in dementia care, suggesting virtual environments can enhance confidence and relationships in long-term care settings.
Rose et al., \cite{b56} described a study where people with dementia were invited to use a HMD-VR in a hospital room with a familiar caregiver in the evaluation. The use of a VR Headset (Samsung Gear VR with a Samsung Galaxy S6 mobile phone (HMD-VR)), allows the user to be fully immersed by controlling the viewing direction by turning their head as they would in the physical world. Users were presented with a ‘menu' of five different VEs from which to choose that include a forest, countryside, sandy beach, rocky beach, and a cathedral. Dementia patients were given a maximum of 15 minutes of HMD-VR exposure, with the VE (s) being viewed through the headset. Data was collected over a two-month period using a mixed techniques methodology that included observations and semi-structured interviews. People with dementia were excited to use HMD-VR and participated effectively in the sessions, according to caregivers. They also stated that HMD-VR had a beneficial impact on people's post-session well-being and encouraged them to do outdoor activities. The careers said that they learned new skills and interests about the person with dementia as a result of attending the session, and that the success of the session made them reassess their involvement in other activities. The sample size was small, and the study was limited to a particular hospital setting, which limited the generalizability of the findings. Despite this, significant data has emerged from this study in relation to the practicality of using HMD-VR technology with a potentially difficult patient group. 
Matsangidou et al., \cite{b57} explored challenges in designing and implementing a VR system for physical training in moderate to severe dementia patients. Their study, conducted in a mental health facility, involved a participatory design process with input from patients and health specialists. Two types of interactions and three exercise types were tested in VR. Study 1 involved a workshop to select appropriate VR content, choosing a forest backdrop and three exercises: climbing a rope, climbing a wall, and seated cable row. Study 2 assessed usability of VR interactions for a boxing exercise. Study 3 allowed patients to use the VR system in a familiar hospital room with their physiotherapist and psychologist present, focusing on task performance, reaction time, and patient independence. Findings suggest the potential of VR physical training for dementia patients and offer guidelines: providing regular feedback, ensuring stable and visible visual targets, tailoring virtual worlds to patient interests, and effectively transitioning patients back to reality to minimize anxiety and confusion. It is thought that encouraging people with generalized anxiety disorder to perform aerobic exercise as a stress-reduction strategy can be accomplished through virtual exercise therapy\cite{b58}. 
\section{Future Perspectives - Designing Elderly Friendly XR Applications}
Immersive technologies, including VR and AR, are rapidly gaining attention as tools that can enhance the lives of elderly individuals, especially as developers increasingly focus on inclusivity and ethical considerations in their design. These technologies can be tailored to support seniors in managing physical health and cognitive decline, with VR-based rehabilitation programs helping older adults recover mobility and regain independence after strokes or injuries \cite{b59}, \cite{b60}. Additionally, VR cognitive training platforms are emerging as promising interventions for managing neurodegenerative diseases like Alzheimer's, providing engaging exercises that stimulate memory retention and cognitive function \cite{b61}. 
Furthermore, AR applications designed for daily life assist seniors with tasks such as navigation, medication management, and an emergency response, which helps them live more independently while ensuring safety \cite{b62}. As developers integrate these technologies into healthcare systems, they are also creating opportunities for remote monitoring, making it easier for healthcare professionals to track patients' progress and adjust treatments \cite{b63}.
As immersive technologies continue to evolve, they also offer significant potential for improving social well-being and reducing isolation among elderly populations. Virtual spaces allow older adults to reconnect with loved ones, interact with peers, or engage in group activities, which have been shown to alleviate feelings of loneliness and promote mental well-being \cite{b64},\cite{b65}. Recent research of Anastasiadou et al., \cite{b66} has created an application aimed at the elderly. Specifically, through the application, the elderly can tell stories from their youth in a pleasant and user-friendly virtual environment designed for this specific target group and help the user to retrieve information that may fade over time while being in an accessible environment.
Moreover, the ability to create virtual environments that simulate past experiences or familiar places offers therapeutic benefits, particularly in reminiscence therapy for those with dementia \cite{b67}. The research of Yondjo and Siette \cite{b68} gathers information from a variety of healthcare providers e.g. general practitioners, nurses and psychologists, involved in the diagnosis and management of dementia to assess how they perceive the applicability, advantages and challenges of using VR for dementia diagnosis. It also explores their views on the feasibility of integrating VR-based tools into routine primary care practices.
The study concludes that while there is enthusiasm about the potential of VR as a diagnostic tool for dementia, its adoption into primary care practice will require addressing both technical and practical challenges, such as accessibility, cost, and healthcare professional training.
However, as developers continue to focus on creating inclusive and accessible experiences, ethical concerns around data privacy, consent, and the emotional impact of virtual environments will need to be addressed to ensure these technologies serve the elderly population responsibly and effectively \cite{b69}. By emphasizing user-centered design, these technologies can help older adults live healthier, more connected lives, enhancing both their physical and emotional well-being.
\section{Conclusions}
As VR technology continues to evolve, the focus is shifting towards designing applications that are more accessible and comfortable for older users. A critical aspect of this evolution is improving VR hardware. Current VR headsets, often bulky and uncomfortable, pose a significant challenge for older users, particularly those with limited dexterity or mobility. Future VR designs are expected to prioritize ergonomics, with lighter, more comfortable headsets that can accommodate the physical needs of older people \cite{b70}. In addition, wearable devices that provide adaptive support—such as those that adjust based on the user’s posture or health data—could further personalize the VR experience, making it more suitable for older people. 
Along with hardware improvements, software development is crucial to creating more age-friendly VR applications. User interface design should take into account the potential cognitive, sensory, and physical limitations of older adults. For example, interfaces should be intuitive and simple, with clear visual and auditory cues. Customizable settings, such as adjustable font sizes and contrast levels, could assist visually impaired users, while voice commands and gesture-based gestures could offer hands-free navigation for people with limited mobility or dexterity \cite{b71}. In addition, VR applications should include features that minimize cognitive overload and provide gentle guidance to help users navigate the virtual environment. Tailored programs, such as cognitive training exercises or social simulations, could be developed specifically to address mental health challenges faced by older adults, such as memory loss or loneliness. 
Another key element for the future of age-friendly XR applications is the integration of social interaction features. Social isolation is a significant concern among older adults, and extended reality offers a unique opportunity to bridge this gap by creating virtual spaces for connection. Future XR platforms may incorporate video calling, virtual meeting spaces, and multi-user environments where older users can interact with family members, friends, and other older adults. These social XR environments could mimic real-life activities, such as virtual games, group exercises, or even community gatherings, helping to enhance a sense of sociability \cite{b72}. As these social features become more prevalent, older users may experience a reduction in feelings of isolation, promoting emotional well-being and engagement. 
In conclusion, the future of XR for older users holds great promise, especially as designers and developers begin to focus on creating accessible, inclusive, and engaging experiences for this demographic. By addressing the specific physical, cognitive, and social needs of older adults, XR applications can become powerful tools for promoting healthy aging. As technology continues to advance, the integration of adaptive systems, improved hardware, and user-centered design will make XR a vital part of older people’s lives, enabling them to maintain their independence, improve their quality of life, and stay connected to others \cite{b73}.

\bibliographystyle{unsrt}

\end{document}